\documentclass{aa}
\usepackage{graphicx}
\begin{document}
\title{V4641Sgr -- Super-Eddington source enshrouded by an extended envelope}

\author{M.Revnivtsev\inst{1,2}, R. Sunyaev\inst{1,2}, M.Gilfanov\inst{1,2}, E. Churazov\inst{1,2}}
\institute{Max-Planck Institut f\"ur Astrophysik,
Karl-Schwarzschild-Str. 1, 85748 Garching, Germany
\and
Space Research Institute, Russian Academy of Sciences,
Profsoyuznaya 84/32, 117810 Moscow, Russia,
}

\authorrunning{Revnivtsev et al.}

\date{Received ; accepted }

\abstract{
An optical spectroscopy  of an unusual fast transient V4641 Sgr
constrains its mass to 8.7-11.7$M_{\odot}$ ($9.6M_{\sun}$ is the best fit
value) and  the distance to 7.4--12.3 kpc (\cite{orosz}). At this distance
the peak flux of 12 Crab in the 2--12 keV energy band, measured by ASM/RXTE,
implies the X-ray luminosity exceeding $2-3\cdot 10^{39}$ erg/s, i.e.
near or above the Eddington limit for a $9.6M_{\sun}$ black hole. 
An optical photometry shows that at the peak of the optical outburst the visual
magnitude increased by $\Delta m_V\ga 4.7^{\rm m}$ relative to the quiescent
level and reached $m_V\la 8.8^{\rm m}$. An assumption that this optical
emission is due to irradiated surface of an accretion disk or a companion
star with the the black body shape of the spectrum would mean that the
bolometric luminosity of the system exceeds $\ga3\cdot 10^{41}~{\rm erg/s}\sim 300
L_{\rm Edd}$. \\
We argue that the optical data strongly suggest presence of 
an extended envelope surrounding the source which absorbs primary X--rays 
flux and reemits it in optical and UV. The data also suggests that this 
envelope should be optically thin in UV, EUV and soft X-rays. 
The observed properties of V4641 Sgr at the  
peak of an optical flare are very similar to those of SS433. This envelope
is likely the result of near or super Eddington rate of mass accretion
onto the black hole and it vanishes during subsequent evolution of the
source when apparent luminosity drops well below the Eddington value. Thus this
transient source provides us direct proof of the dramatic change in the
character of an accretion flow at the mass accretion rate near or above the
critical Eddington value as predicted long time ago by the theoretical models.
}

\maketitle

\sloppypar
\section{Introduction}

The X-ray transient V4641~Sgr was discovered in February 1999 
(\cite{sax_discovery}, \cite{rxte_discovery}) and since
then it has demonstrated moderate X-ray activity during approximately a 
half of a year at a level of $\sim$10 mCrab (\cite{markwardt_pca},
 \cite{mikej4641}). In September 1999 it entered a period
of violent X-ray and optical activity, during which the X-ray flux peaked 
at $\sim12$ Crab (\cite{asm_flare}) and observed optical brightness peaked at
$m_V\sim8.8$ (\cite{stubbings99}, \cite{kato99}). After a few days of
a strong activity in the optical (\cite{kato99}),
 radio (\cite{hjellming00}) and X-ray (\cite{mikej4641}) bands the source 
became undetectable in radio and X-rays, and returned back to the quiescent
level in optics. It should be noted that an unusual optical activity of this 
object was discovered more than 20 years ago by Goranskij (1978), indicating 
that the source is a recurrent transient.

A detailed analysis of the X--ray data will be reported elsewhere
(Revnivtsev et al., 2001). In this Letter we concentrate on the implications
of the available optical data (\cite{gor78}, \cite{gor90}, \cite{kato99}, 
\cite{orosz}) and demonstrate that these data strongly suggest presence of
an extended envelope surrounding the source.

\section{Optical evidence for the envelope}
Optical spectroscopic observations of V4641 Sgr performed at the
quiescence led to its identification as a high mass black 
hole binary with an orbital period of $\approx 2.81$ days and the primary
and secondary masses of $\approx 9.6M_{\sun}$ and $\approx
6.5M_{\sun}$.  Optical observations also constrains the
source distance to 7.4--12.3 kpc (\cite{orosz}). This new distance
measurements dramatically changed our understanding of the source. At this
distance the peak flux of 12 Crab  measured by ASM/RXTE implies that the
2-12 keV luminosity of the source exceeds $2-3\cdot 10^{39}$ erg/s, i.e. is
near or above the Eddington luminosity for a $9.6M_{\sun}$ black hole.

\begin{figure}
\centering
\includegraphics[width=\columnwidth,bb=18 172 590 580,clip]{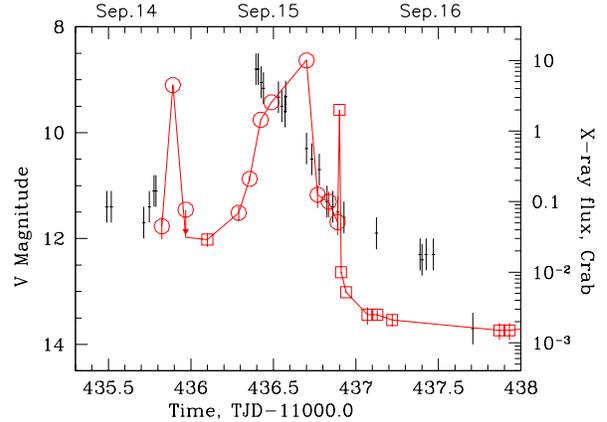}
   \caption{The light curves of V4641 Sgr in the optical V-band
(crosses, from \cite{kato99}) and and in X-rays near the peak of main outburst in Sep.1999.
 The RXTE/ASM points (1.5-12 keV) are shown by 
open circles, the RXTE/PCA (3--20 keV) - by open squares (\cite{mikej4641}).
           }
      \label{fig:lc_zoom}
\end{figure}

The most remarkable property of V4641 Sgr was revealed by the optical
photometry during the giant outburst in Sep. 1999. The available optical and
X--ray data are shown in Fig.\ref{fig:lc_zoom}. The whole outburst was
rather short and there is a gap in optical data between Sept. 14.8
and 15.4. The X-ray data are mostly those of ASM/RXTE with the sampling
interval of approximately 1.5 hour. The existing optical data suggest that
the optical brightness started to increase around Sept. 14.8, 1999,
approximately at the time of the first outburst in X-rays. The peak of the
optical brightness however does not coincide with
any of the 3 peaks observed in X-rays. Moreover, during the peak of X-ray flux
on $\sim$Sep.15.7, 1999 the optical flux shows steady decline
uncorrelated with the behavior in X-rays. In general the data shown in
Fig.\ref{fig:lc_zoom} suggest a relatively smooth
evolution of the optical flux and an erratic multi-peaked behavior in
the X-rays. The peak in the optical flux appears to coincide with the
minimum of X-ray flux although the possibility of  short outburst(s)
occurring between the X-ray points can not be entirely excluded.

A lower limit on the peak source luminosity can be obtained from the ASM
data. The maximum flux measured by ASM during the giant outburst was
$\approx 12.1$ Crab. The ASM observation, likely missed the very peak of
the outburst, therefore the actual peak flux might be higher. Using the
source count rates in three ASM energy channels and the Crab count rates 
in these channels as a reference for the flux estimates,
we can obtain the 1.3-12 keV energy flux from V4641 Sgr at the peak
of the X-ray outburst  --  $F_{\rm x}\ga 3.8\cdot 10^{-7}$ erg/s/cm$^2$. At the
distance of 9.6 kpc (7.4 -- 12.3 kpc at the 90\% confidence) it would
correspond to the 1.3-12 keV luminosity in excess of $ 4.2\cdot
10^{39}$ ($2.5-6.9\cdot 10^{39}$) erg/s. On the other hand, the
Eddington luminosity for a $9.6M_{\sun}$ ($8.7-11.7M_{\sun}$)
 is $\sim1.3\cdot 10^{39}$ erg/s ($1.2-1.6\cdot 10^{39}$
erg/s).  We therefore conclude that at the peak
of the outburst of V4641 Sgr the 1.2-12 keV luminosity {\em alone} exceeded
critical Eddington luminosity by a factor of a few: $\ga 3.2L_{Edd}$
($(1.6-5.7)  L_{Edd}$).

Because of the gap in data between Sept. 14.8 and 15.3 the optical
observations  also likely missed the peak of the optical 
outburst, therefore the maximal visual brightness at the peak corresponds
to $m_V \la 8.8^{\rm m}$. 
The interstellar reddening in the direction of the source, 
${\rm E(B-V)}=0.32\pm 0.1$ (\cite{orosz}), corresponds to the extinction of 
${\rm A_V}\approx 3.2{\rm E(B-V)} \approx 1.0\pm 0.3$. The extinction
corrected flux at $\lambda\sim0.55~ \mu{\rm m}$ exceeds $F_{\nu}\ga 6.7
\cdot 10^{-6}$ erg/s/cm$^2$/keV, which approximately corresponds to the
source luminosity in the optical spectral band of $L_{\rm opt}\sim10^{38.2}$
ergs/s. Note here, that even the optical data {\em alone} give us the
near-Eddington source luminosity.

Increase of optical flux is observed during X--ray outbursts in many X-ray
binaries (e.g. \cite{vpar}). The star itself can not change its internal brightness
significantly in a time scale of days or months. Therefore it is usually
assumed that the optical emission in the low mass X-ray binaries (which 
have very weak optical companions) is due to the
irradiated accretion disk and the irradiated side of the secondary star (e.g.
Lyutyi \& Sunyaev, 1976).
However, in the case of high mass companions (like V4641~Sgr), which already
have quite powerful optical star companion, it is extremely difficult to
increase the optical luminosity of the system by a factor of $\sim$100. To
the best of our knowledge, this is the only case, when the high mass binary
system optical brightness has changed by such a large factor. If we assume
that the peak optical flux is due to the optically thick thermal emission
of the accretion disk or a companion star we can estimate the bolometric
luminosity of system required to explain the observed optical flux. 

\begin{figure}
\centering
\includegraphics[width=\columnwidth,bb=18 163 590 650,clip]{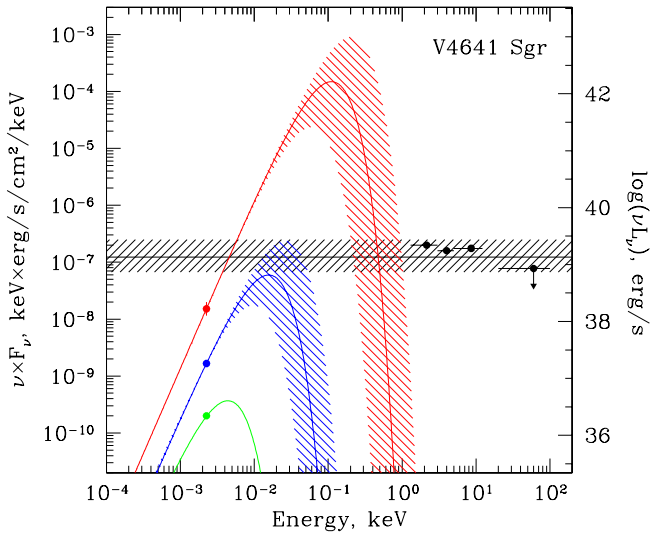}
   \caption{The approximation of the spectra of V4641 Sgr at the different stages 
of the outburst by the optically thick emission model (black body) 
using the optical points ($\lambda\sim0.55\mu$m, $h\nu\sim2$eV).
Upper curve corresponds to maximal observed optical brightness. 
In the X-ray band (1--100 keV) ASM/RXTE and BATSE/CGRO points are shown 
(not simultaneous with optical points). The shaded area 
around the curves show the allowed region, using some freedom in the system
 parameters. Horizontal shaded area shows the Eddington level for the black 
hole with mass $M_{\rm BH}\sim10 M_{\odot}$ using different source distance in the range 7.4-12.3
 kpc.
           }
      \label{fig:spe}
\end{figure}

The size of the accretion disk $R$ in the
binary system can be assumed to be a fraction $k$ of the size of the Roche
lobe of the primary, where $k$, according to different estimations, is of
the order of  0.3--0.6 (see e.g \cite{paczynski77}). Using this radius of
the emitting  region and the observed optical flux we can estimate the
temperature of the emitting region assuming the simplest Plank 
shape of the spectrum. The obtained temperature is close to
$T\sim3\cdot10^{5}$K. Since the irradiated disk is optically thick and is
emitting black body spectrum its bolometric luminosity can be estimated as
$2\pi R^2 \sigma T^4 \sim$few$\cdot 10^{41}$ ergs/s. In
Fig.\ref{fig:spe} we present the black body models 
of the spectra of V4641~Sgr during three stages of its flux history - from
the bottom to the top - the quiescent, pre-outburst and the peak optical
activity. The shaded areas around the spectra show the allowed regions of
spectral flux densities, taking into account the freedom in the system
parameters. It is immediately seen that within the irradiation model the
bolometric luminosity of V4641~Sgr at the peak of optical outburst should of
the order of $10^{41}$--$10^{43}$ ergs/s for any choice of the binary system 
parameters.  The same conclusion is reached if we assume that the irradiated
surface of the secondary star is responsible for the observed optical flux. 

It is very unlikely that the source powered by an accretion can have
the luminosity 2--3 orders of magnitude higher than the Eddington
value. Therefore one has to consider alternative scenarios.

As was suggested already in 1970s (e.g. \cite{ss73}), a super-critical
accretion onto a black hole could result in the formation of a wind, or 
geometrically thick envelope. Such an envelope, under certain circumstances,
could be optically thick for 
{\em a)} X-ray photoabsorption, and {\em b)}
infrared and optical emission, but still  optically thin in UV, EUV and 
soft X-rays, where the bulk of energy can be emitted. Below we assume that
such an envelope, intercepting X--ray flux from the source and reradiating
in optical and UV bands, is responsible for the observed optical emission.
Detailed calculations of the radiation transport in the envelope is
beyond the scope of our paper, but some rough estimates of the envelope
temperature and density are possible. 

First of all, let us assume that the optical emission has a thermal origin.
If one takes too low temperature of the emitting region, e.g. $T<3\cdot 
10^4$K, then, given the observed optical flux of the system, 
the required size of the emitting region (even under assumption of a black
body spectrum) is much larger than the size of the
binary  system, which is unlikely. On the other hand, for a too high
temperature of the envelope, e.g. $T>3\cdot 10^6$K, the required
size of the emitting region is smaller, but the bolometric luminosity
inevitably will exceed the Eddington value even under assumption of an optically
thin emission spectrum. Only for some intermediate values of temperature the
size of the emitting region can be made comparable to the binary system size
with the bolometric source luminosity below the Eddington value. In the
discussion below we will assume the temperature of the envelope of
$T\sim3\cdot 10^5$K. The envelope is then optically thick in optics, is
getting optically thin in UV and EUV and is optically thick 
for X-ray photoabsorption.

In order to absorb (via photoabsorption) most of the energy emitted in
X-rays  the hydrogen column density of the envelope  $\int{ndl}$ (where $n$
is  the hydrogen density of the envelope) has to be at least of the order of
$10^{24-25}$cm$^{-2}$, which means Compton thick envelope. On the other hand, 
maximum radiation efficiency in the optical band is achieved if the the line
of sight emission measure of the envelope is $\int{n^2dl} \ge
10^{37-38}$cm$^{-5}$ (depending on the contribution of the opacity in
the lines to the total opacity of the envelope). Such line of sight emission
measure implies that the envelope is optically thick in the optical band
(for the assumed temperature of $T\sim3\cdot 10^5$K).
Line of sight emission measure however should not be much larger than this
value. Otherwise the envelope would become optically thick in UV and EUV the
the luminosity would exceed the Eddington value. Thus, assuming the
homogeneous envelope (which is obviously large oversimplification) we can
conclude that for the size of the envelope $R\sim$few$\cdot R_{\odot}$ (i.e.
of the order of size of the Roche lobe of the primary) the density 
 should be about $n\sim10^{12-14}$cm$^{-3}$. The total mass of the envelope
is then $M_{\rm env}\sim\int{ndV}\sim10^{23-25}$g. It is interesting that
for with an Eddington mass accretion rate of $\dot{M}_{\rm
Edd}\sim10^{19}$g/s (for a $M\sim10M_{\odot}$ black hole) such a mass can be
accumulated during $10^4$--$10^6$ s. Cooling time of the envelope is much
shorter -- of the order of seconds.

Thus the optical data suggest that there is an absorbing/reprocessing envelope
surrounding the source. In such a model a 
relatively smooth behavior of the optical luminosity would reflect the total 
intrinsic luminosity of the source and the amount of X--rays absorbed by 
the reprocessing region. The evolution of $L_{\rm opt}/L_{\rm x}$
 suggests that the geometry of absorbing/reprocessing region changes with
time. At the maximum of the  optical light curve most of the X--rays are
probably absorbed, while later  
(at the X--ray maximum) the dominant fraction of X--rays is directly observed.
Rapid changes in X--ray flux could be due to changes of the geometry of the 
absorbing region (e.g. edge of the torus obscuring line of sight) or thermal 
instability in the gas which causes fragmentation of the medium into separate 
clouds. During these transitions the distribution of the energy emitted in
different energy bands (e.g. in optical/UV and in X-rays) could change
strongly depending on the parameters of the envelope, while the bolometric
luminosity of the source behaves much smoothly and reflects the mass
accretion rate onto the black hole.

 Note that the enhanced optical activity,
observed during approximately 2 weeks before the giant outburst (the middle
curve on the Fig.2, \cite{kato99}) and the optical activity observed by 
Goranskij (1978) could also be explained in the frame of
the model with the emitting extended envelope. During these observations 
$m_{\rm V}$ was $\sim12^m$ (two magnitudes brighter than that during 
the quiescence). Assuming a black body shape of the 
spectrum and the size of the emitting region of the order of the size of the 
accretion disk one can estimate the temperature of the emitting region
to be of the order of
 40,000-50,000K. This temperature is again far too high for the irradiated
disk or a star (e.g Lyutyi \& Sunyaev 1976), especially taking 
into account the value of the X-ray flux, observed simultaneously (\cite{mikej4641}).
The observed optical flux at this stage could still correspond to 
near-Eddington bolometric luminosity if the temperature of the envelope is
 10 times lower than that during the brightest optical flare. This is 
possible if the line of sight emission measure and geometrical dimensions 
of the envelope are the same. The optical depth for UV and EUV radiation 
in this case would be much higher. 
 
\section{Comparison with SS 433}
The difficulties with the estimates of the bolometric luminosity of 
V4641 Sgr during the outburst described above resemble those with the well
known  source SS 433. Hubble Space Telescope observations of SS 433 
have detected the source in UV at $\sim1500\dot{A}$. The brightness  
temperature of the optical emission was estimated to be $T\sim72,000$K
(\cite{hst_ss433}). The energy flux, observed in UV spectral band is 
$\sim2\cdot10^{-16}$ergs/s/cm$^2/\dot{A}$. After the correction for the
 interstellar absorption, the luminosity of the source in UV is 
$L\sim 10^{39}$ergs/s. The bolometric source luminosity, calculated using 
the parameters of the optical-UV emission, presented in Dolan et al (1997)
is around $L_{\rm bol}\sim1.3\cdot10^{40}$ ergs/s. 
 This value is strongly super-Eddington even for mass of the $10M_{\odot}$ 
compact object.
However, if we assume that in the vase of SS 433 we also observe the envelope, 
which is optically thick in optical and UV spectral bands, but still 
optically thin in EUV spectral band, then the estimate of the 
source bolometric luminosity can be strongly reduced, and become 
close to Eddington luminosity for the $10M_{\odot}$ black hole. 

\section{Conclusion}
In this Letter we showed that the optical observations of X-ray transient
V4641~Sgr strongly suggest that there is a massive envelope enshrouding the
source. This envelope is likely the result of near or super Eddington rate
of mass accretion onto the black hole. During this state the V4641~Sgr
closely resembles another unique object -- SS 433, which is also thought to
be a Super Eddington accretor. Fraction of bolometric luminosity emitted in
different energy bands strongly depends on the parameters of the envelope and, thus, observed optical and X--ray fluxes can evolve in a complicated and
uncorrelated way. As the mass accretion rate onto the black hole drops well
below the Eddington value the envelope vanishes. Thus this transient source
provides us for the first time a direct proof of the dramatic change in the
character of an accretion flow at the mass accretion rate near or above the
critical Eddington value as it was predicted long time ago by the theoretical models.

\begin{acknowledgements}
This research has made use of data obtained through the High Energy
Astrophysics Science Archive Research Center Online Service, provided 
by the NASA/Goddard Space Flight Center. 
The work was done in the context of the research network
"Accretion onto black holes, compact objects and protostars"
(TMR Grant ERB-FMRX-CT98-0195 of the European Commission).

\end{acknowledgements}

\end{document}